\documentclass[prl,twocolumn,showpacs,preprintnumbers,amsmath,amssymb,superscriptaddress]{revtex4-1}

\usepackage{graphicx}
\usepackage{dcolumn}
\usepackage{bm}
\usepackage{color}

\begin{document}

\title{Irreversibilities and efficiency at maximum power of heat engines: illustrative case of a thermoelectric generator}

\author{Y. Apertet}\email{yann.apertet@u-psud.fr}
\affiliation{Institut d'Electronique Fondamentale, Universit\'e Paris-Sud, CNRS, UMR 8622, F-91405 Orsay, France}
\author{H. Ouerdane}
\affiliation{CNRT Mat\'eriaux UMS CNRS 3318, 6 Boulevard Mar\'echal Juin, 14050 Caen Cedex, France}
\author{C. Goupil}
\affiliation{Laboratoire CRISMAT, UMR 6508 CNRS, ENSICAEN et Universit\'e de Caen Basse Normandie, 6 Boulevard Mar\'echal Juin, F-14050 Caen, France}
\author{Ph. Lecoeur}
\affiliation{Institut d'Electronique Fondamentale, Universit\'e Paris-Sud, CNRS, UMR 8622, F-91405 Orsay, France}

\date{\today}

\begin{abstract}
Energy conversion efficiency at maximum output power, which embodies the essential characteristics of heat engines, is the main focus of the present work. The so-called Curzon and Ahlborn efficiency, $\eta_{\rm CA}$, is commonly believed to be an absolute reference for real heat engines; however a different but general expression for the case of stochastic heat engines, $\eta_{\rm SS}$, was recently found and then extended to low-dissipation engines. The discrepancy between $\eta_{\rm CA}$ and $\eta_{\rm SS}$, unexplained so far, is here analyzed considering different irreversibility sources of heat engines, of both internal and external types. To this end, we choose a thermoelectric generator operating in the strong coupling regime as a physical system to qualitatively and quantitatively study  the impact of the nature of irreversibility on the efficiency at maximum output power. In the limit of pure external dissipation, we obtain $\eta_{\rm CA}$, while $\eta_{\rm SS}$ corresponds to the case of pure internal dissipation. A continuous transition between from one extreme to the other, which may be operated by tuning the different sources of irreversibility, also is evidenced.
\end{abstract}

\pacs{05.70.Ln, 84.60.Rb}
\keywords{Finite-time thermodynamics, thermoelectric energy conversion, irreversibilities}

\maketitle

\paragraph*{Introduction}

When Sadi Carnot was reflecting on the motive power of heat, the first law of thermodynamics was not yet formulated. Though the so-called caloric theory, which states that heat can neither be created nor destroyed, was widely accepted, Carnot proposed an idealized model of heat engine showing that the fraction of energy that can be extracted as work from heat transiting between two thermostats at temperatures $T_{\rm hot}$ and $T_{\rm cold}$ respectively, cannot exceed $1-T_{\rm cold}/T_{\rm hot}$, the upper limit which defines the Carnot efficiency $\eta_{\rm C}$ \cite{Carnot1824}. This limit can be reached only if the process is fully reversible. Carnot's basic assumption that heat is conserved, is incorrect but his intuition paved the way to the second law of thermodynamics and the related concept of irreversibility.

A reversible transformation in a thermodynamic system is \emph{quasi-static}, and hence requires an infinite time to complete. As a consequence the ideal Carnot engine is a \emph{zero-power} engine; further, it is off the arrow of time since no dissipative element ensures causality. For practical purposes, real thermodynamic engines must produce power and not just work to be useful; so one usually seeks maximum efficiency at nonzero power or, even more, maximum output power. Causality can be restored by introducing dissipation through finite thermal conductances between the ideal Carnot engine and the heat reservoirs, as Chambadal \cite{Chambadal1957}, Novikov \cite{Novikov1958}, and Curzon and Ahlborn~\cite{Curzon1975} did to derive a simple, yet general, expression for the efficiency at maximum power: 

\begin{equation}\label{effCA}
\eta_{_{P_{\rm max}}}= 1-\sqrt{T_{\rm cold}/T_{\rm hot}}\equiv\eta_{\rm CA}
\end{equation}

\noindent known as the Curzon-Ahlborn efficiency. These seminal works put forward a then new kind of system termed as \emph{endoreversible} (reversible only when considered alone, but not when finite thermal contacts are involved~\cite{Rubin1979}) and gave rise to finite time thermodynamics. Equation \eqref{effCA} was rederived as a general result of linear irreversible thermodynamics in Ref.~\cite{VandenBroeck2005}.

Recently, another general yet different expression for efficiency at maximum power: 

\begin{equation}\label{effSS}
\eta_{_{P_{\rm max}}}=\eta_{\rm C}/(2-\gamma \eta_{\rm C})\equiv \eta_{\rm SS},
\end{equation} 

\noindent where $\gamma$ is a parameter related to the ratio of entropy production at each end of the engine, was obtained by Schmiedl and Seifert using a stochastic heat engine model~\cite{Schmiedl2008}. More recently, an extension of this result to the class of low-dissipation heat engines was reported by Esposito and co-workers~\cite{Esposito2009,Esposito2010}. The main purpose of this article thus is to explain and discuss the discrepancy between $\eta_{\rm CA}$ and $\eta_{\rm SS}$.

We observed that the hypothesis used in Ref.~\cite{Schmiedl2008} to obtain $\eta_{\rm SS}$ differs from the assumption of the endoreversible engine in that no dissipative thermal contacts are involved, and irreversibilities arise only from internal processes (this class of engine is referred to as an \emph{exoreversible} engine~\cite{Chen1997}). This observation led us to focus on sources of irreversibility in real thermal engines. These sources are varied and include friction and heat leaks. In models which account for the coupling of the engine to the reservoirs, the irreversibility may also originate in the finiteness of the heat transfer rate. A continued increase of the speed of a heat engine operation results in a decrease of both power and efficiency because of friction and finite-rate heat transfer; conversely, in a slow regime operation heat leaks become the preponderant irreversibility source which negatively impacts on output power and hinders efficiency at finite-rate heat transfer.

Thermoelectric generators (TEG) are devices which couple electric and heat currents, and hence constitute a very interesting type of real thermal engines, for which three sources of irreversibilities are identified: the Joule effect, obviously an internal process, the heat leak represented by the open-circuit thermal conductance $K_{0}$, and the dissipative thermal contacts to the heat reservoirs. Twenty years ago, Gordon \cite{Gordon1991} studied the impact of these three kinds of irreversibility on the behavior of a TEG comparing the relation between the produced power $P$ and the efficiency $\eta$ for various cases. Interestingly he demonstrated that a TEG with only Joule dissipation or only disspative thermal contacts (endoreversible case) exhibits the same behavior: an open $P$ vs. $\eta$ curve where the electrical open circuit condition allows to reach the Carnot efficiency. On the contrary when heat leaks are introduced, the $P$ vs. $\eta$ curve becomes closed: in open-circuit condition the efficiency vanishes as for closed-circuit condition. These results show that heat leaks should not be treated on the same footing as Joule heating and thermal contact dissipation. We emphasize that the case where only heat leaks are considered is unphysical and presents no interest \emph{per se} because electrical transport is then not allowed to take place. This may explains why this situation is not treated in Ref.~\cite{Gordon1991}.

This problem can be compared to that of the connection of a perfect capacitor to a perfect voltage generator: since such capacitor cannot sustain a potential discontinuity, it is impossible to connect both components for a practical purpose unless a dissipative element, such as a resistor, is introduced in the circuit. But one cannot place this irreversibility source at random: the dissipation is useful to realize the coupling if the resistor is connected in series with the generator but useless if placed in parallel. In that case, as for the thermal conductance for the TEG, dissipation occurs without resolving the causality issue. In this article, we thus analyze a model TEG that presents no heat leaks. This assumption is equivalent to that of strong coupling defined by Van den Broeck \cite{VandenBroeck2005} as the heat flux is thus only composed of an advective term \cite{Apertet2011a} and hence proportionnel to the electron flux. We deal with two irreversibility sources only: an internal one (Joule heating) and an external one (dissipative thermal coupling).

In this context, a question naturally arises: How internal and external irreversibilities can be compared? We recently demonstrated the importance of thermal contacts in practical applications such as a thermoelectric generator coupled to heat reservoirs with nonideal heat exchangers \cite{Apertet2011b}. Indeed, the impact of the thermal contacts on the electrical properties of the TEG is such that an additional electrical resistance appears in the basic Th\'evenin model of the TEG. The comparison between this additional electrical resistance and the standard Th\'evenin internal electrical resistance provides a means to quantify the internal and external sources of irreversibility in the system.

In this article we build on the works of Onsager~\cite{Onsager1,*Onsager2}, Callen~\cite{Callen1} and Domenicali~\cite{Domenicali1} on irreversible processes to study a generic model of thermoelectric generator connected to two temperature reservoirs. This framework permits a very efficient and physically transparent description of the coupling of the laws of Ohm and Fourier, which govern the properties of thermoelectric generators. Considering two limit cases for irreversibility sources: that of pure external irreversibility, and that of pure internal irreversibility, we find that the efficiency at maximum power in these two extremal cases corresponds exactly in one case to the Curzon-Ahlborn efficiency, in the other to that calculated by Schmield and Seifert~\cite{Schmiedl2008}. We also derive an analytic expression of the efficiency when the conductances of the thermal contacts placed at both ends of the thermoelectric module are equal, and we discuss the obtained expression in the light of previously published results. This discussion is then extended to the dissymetric case.\\

\paragraph*{Thermoelectric model}

\begin{figure}
	\centering
		\includegraphics[width=0.43\textwidth]{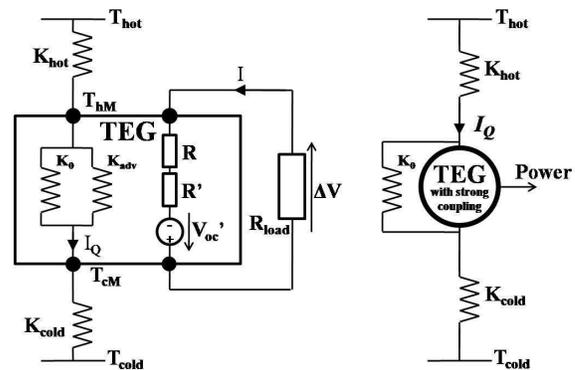}
	\caption{Thermoelectric (left) and thermodynamic pictures of the thermoelectric generator.}
	\label{fig:figure1}
\end{figure}

We consider a model thermoelectric generator connected to two temperature reservoirs, depicted in Fig.~\ref{fig:figure1}. The temperatures of the heat reservoirs are $T_{\rm cold}$ and $T_{\rm hot}$ respectively. The thermal contacts are characterized by two thermal conductances $K_{\rm cold}$ and $K_{\rm hot}$ so that the total contact thermal conductance is given by $K_{\rm contact}=K_{\rm cold}K_{\rm hot}/(K_{\rm cold}+K_{\rm hot})$. The TEG is characterized by its isothermal electrical resistance $R$, its Seebeck coefficient $\alpha$, both constant inside the module, and its thermal conductance $K_{\rm TEG}$, which is composed of a conductive part $K_0$ associated with heat leaks and an advective part $K_{\rm adv}$ associated with the electrical current \cite{Apertet2011a}. In the strong coupling regime the isothermal conductance of the generator is supposed to be zero and so the average heat flux $I_{Q}$ is proportionnal to the electrical current $I$ \cite{VandenBroeck2005}. In the electrical circuit a resistance $R'$ is due to the presence of the finite thermal contacts as demonstrated in Ref.~\cite{Apertet2011b}; under the strong coupling assumption it is given by $R'=\alpha^2T'/K_{\rm contact}$, $T'$ being the average temperature inside the TEG. The voltage across the generator, representing the thermoelectric conversion, is $V_{\rm oc}'=\alpha(T_{\rm hot}-T_{\rm cold})$.

The temperatures at both ends of the thermoelectric module, $T_{\rm hM}$ and $T_{\rm cM}$ are explicitely given by \cite{Yan1993}:

\begin{subequations}
\begin{equation}\label{eq:ThM}
T_{\rm hM}=\frac{K_{\rm hot} T_{\rm hot}+\frac{1}{2}RI^2}{K_{\rm hot}+\alpha I}
\end{equation}
\begin{equation}\label{eq:TcM}
T_{\rm cM}=\frac{K_{\rm cold} T_{\rm cold}+\frac{1}{2}RI^2}{K_{\rm cold}-\alpha I}
\end{equation}
\end{subequations}

\noindent Since the average temperature $T'=(T_{\rm hM}+T_{\rm cM})/2$ depends on the working conditions, the resistance $R'$ does too. To remove this dependence we define a resistance $R''$ given by $R''=\alpha^2T/K_{\rm contact}$ with $T=(T_{\rm hot}+T_{\rm cold})/2$. As a first approximation $R'\approx~R''$. A simple expression for the produced power as a function of $T_{\rm hM}$ and $T_{\rm cM}$ reads:
\begin{equation}\label{eq:P}
P=\alpha (T_{\rm hM}-T_{\rm cM})I - R I^2
\end{equation}

\noindent The full analytic expression of the output power $P$ as function of the electrical current $I$ is cumbersome, and can be found in Ref.~\cite{Yan1993}. The conversion of the heat current into the electric power thus is characterized by the efficiency $\eta$:
\begin{equation}\label{eq:efficiency}
\eta=\frac{\alpha(T_{\rm hM}-T_{\rm cM})-RI}{\alpha T_{\rm hM}-RI/2}
\end{equation}

\noindent All the quantities involved here depend on the electrical current $I$, so we have to calculate them numerically as a function of $I$ in the generator regime to extract the efficiency at maximum power $\eta_{P_{\rm max}}$ for various values of the TEG internal resistance $R$. All other parameters of the TEG, including the thermal contact conductances, are fixed (their values do not influence the result on $\eta_{P_{\rm max}}$).

\begin{figure}
	\centering
		\includegraphics[width=0.43\textwidth]{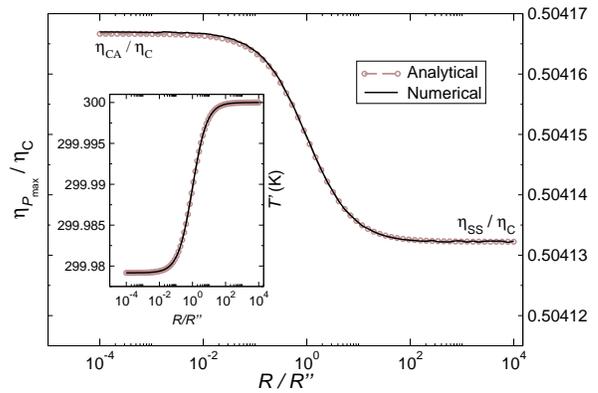}
	\caption{Efficiency at maximum power scaled to the Carnot efficiency and mean temperature $T'$ (inset) as functions of the ratio $R/R''$.}
	\label{fig:figure2}
\end{figure}

We focuse first on the \emph{symmetric} configuration characterized by the equality of the thermal contact conductances: $K_{\rm hot} = K_{\rm cold}$. The efficiency at maximum power, $\eta_{P_{max}}$, is represented as a function of the ratio $R/R''$ for a TEG working between $T_{\rm cold} = $ 295 K and $T_{\rm hot} = $ 305 K, in Fig.~\ref{fig:figure2}. We recover the expected behavior for extremal cases: if the sources of irreversibility are mainly external ($R/R''\rightarrow~0$) then we obtain $\eta_{P_{\rm max}} = \eta_{\rm CA}$, which agrees with the calculation of Ref.~\cite{Agrawal1997}; conversely, if the sources are mainly internal ($R/R''\rightarrow\infty$), we obtain the Schmiedl-Seifert efficiency $\eta_{P_{\rm max}}=\eta_{\rm SS}=\eta_{\rm C}/(2-\eta_{\rm C}/2)$, since $\gamma=1/2$ for this particular heat engine (as discussed in the next page). Furthermore we note a continuous transition between these two limits.\\

\paragraph*{Analytical expression for $\eta_{_{P_{max}}}$}

To gain insight into the dependence of the efficiency at maximum power on the external and internal irreversibilities when $K_{\rm hot}=K_{\rm cold}$, we derive an analytic expression for the efficiency. First, we express the electrical current at maximum power $I_{P_{\rm max}}$ as \cite{Apertet2011a}:
\begin{equation}
\label{eq:IPmax3}
I_{P_{\rm max}}=\frac{\alpha}{2}\frac{(T_{\rm hot}-T_{\rm cold})}{R+R'}
\end{equation}

\noindent Then, to obtain the average temperature $T'=(T_{\rm hM}+T_{\rm cM})/2$ inside the TEG, we make the approximation that $I_{P_{\rm max}} = \alpha(T_{\rm hot}-T_{\rm cold})/2(R+R'')$ in Eqs.~(\ref{eq:ThM}) and (\ref{eq:TcM}). So $T'$ is given by: 

\begin{equation}\label{tppmax}
T'_{P_{\rm max}}=T-\frac{R''}{R''+R}\frac{\Delta T^2}{16T},
\end{equation}

\noindent where $\Delta T=T_{\rm hot}-T_{\rm cold}$. The second term on the right hand side of Eq.~\eqref{tppmax}, leading to a deviation from $T$ in the case of an overwhelming contribution of external dissipation, is important to recover the Curzon-Ahlborn efficiency. This variation of $T'$ reflects a response of the system when the internal dissipation constraint is relaxed, and it is verified by the exact numerical calculation as shown in the insert of Fig.~\ref{fig:figure2}. This dependence of the temperature on the ratio $R/R''$ cannot be ignored so we have to use the above expression of $T'$ in the definition of $R'$ instead of the approximation $T'=T$. Replacing this whole form of $R'$ in Eq.~(\ref{eq:IPmax3}) and using Eq.~(\ref{eq:efficiency}), we obtain the following analytic expression for the efficiency at maximum power in the symmetric configuration:
\begin{eqnarray}\label{eq:efficiencyPmax}
\eta_{P_{\rm max}}^{\rm sym} & = &\frac{\eta_{\rm C}}{2} \frac{1+\frac{\displaystyle \eta_{\rm C}}{\displaystyle 2(2-\eta_{\rm C})} \frac{\displaystyle R''}{\displaystyle R''+R}}{1-\frac{\displaystyle \eta_{\rm C}}{\displaystyle 4}\frac{\displaystyle R}{\displaystyle R''+R}}
\end{eqnarray}

\noindent Only leading terms up to the third order in $\eta_{\rm C}$ were retained for the sake of tractability. For $R/R''\rightarrow \infty$, we recover the expression found by Schmiedl and Seifert \cite{Schmiedl2008}; for $R/R''\rightarrow 0$, an expansion to the third order in $\eta_{\rm C}$ leads to Curzon-Ahlborn efficiency developped to the same order: 
\begin{equation}
\eta_{P_{\rm max}}=\frac{\eta_{\rm C}}{2}+\frac{\eta_{\rm C}^2}{8}+\frac{\eta_{\rm C}^3}{16}+o(\eta_{\rm C}^4)
\end{equation}

\noindent The coefficient at the second order in Carnot efficiency remains the same:  1/8, over the \emph{whole range} of variation of the ratio $R/R''$; this result is in perfect agreement with that of Esposito and co-workers \cite{Esposito2009}. As demonstrated in Fig.~\ref{fig:figure2} this analytical expression reproduces well the curve derived from exact formulas.\\

\paragraph*{Insight into the dissymmetric configuration}
The result of Curzon and Ahlborn is interesting and powerful in that it does not depend on the repartition of the dissipated energy between hot and cold contacts. The dissymmetric configuration is characterized by different values of the termal contact conductances $K_{\rm hot} \neq K_{\rm cold}$. However, $K_{\rm contact}$ is kept constant. 

\begin{figure}
	\centering
		\includegraphics[width=0.43\textwidth]{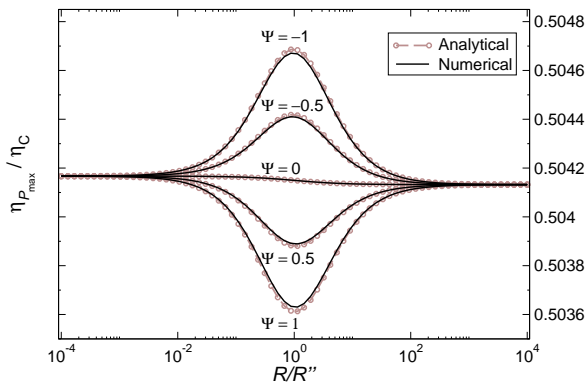}
	\caption{Efficiency at maximum power versus internal electrical resistance scaled to the Carnot efficiency for various values of $\Psi$. Comparison of numerical and analytical results.}
	\label{fig:figure3}
\end{figure}

To proceed in our analysis, we introduce a contrast function $\Psi=(K_{\rm hot}-K_{\rm cold})/(K_{\rm hot}+K_{\rm cold})$ that characterizes the degree of symmetry. Whatever the value of the ratio $R/R''$ is, the numerically computed efficiency at maximum power exhibits a dependence on $\Psi$ that is linear (not shown here). This dependence is enhanced when internal and external dissipations are identical, whereas if one of the irreversibilities vanishes, the efficiency at maximum power remains independent of the degree of symmetry of the thermal conductances. This is coherent with the result of Curzon and Ahlborn for $R/R''\rightarrow0$. For $R/R''\rightarrow\infty$, since internal dissipation is the leading contribution, this behavior can be explained by the intrinsic symmetry of Joule heating: each end of the thermoelectric module receives half of the heat thus produced; this internal symmetry implies that $\gamma$ = 1/2, in the above expression of $\eta_{\rm SS}$ . We cannot explain yet such a dependence for a mixed internal-external contributions of irreversibilities. This clearly is an open question. As of yet, we can only propose an educated guess of the dependence:
\begin{equation}
\eta_{P_{\rm max}}=\eta_{P_{\rm max}}^{\rm sym}+2\Psi\frac{RR''}{(R+R'')^2}~\eta_{\rm C}^3
\end{equation}

\noindent This formula fits well to the numerical result and thus presents an interest: since the dissymmetric configuration has no influence before the third order in the Carnot efficiency is reached, the coefficient 1/8 at the second order still is present even in the dissymmetric configuration. This is sufficient to capture the main features of the influence of the degree of symmetry $\Psi$ on the efficiency at maximum power $\eta_{P_{\rm max}}$ even if we observe a small discrepancy in comparison to the exact result as shown in Fig.~\ref{fig:figure3}: higher orders terms are necessary to obtain a full agreement. In the intermediate situation, where $R$ is comparable to $R''$, the thermal contact with the higher conductance must be placed on the colder side to improve $\eta_{\rm C}$. We do not have yet a satisfactory explanation to propose for this fact. Schmiedl and Seifert~\cite{Schmiedl2008} showed that in the general case of heat engines, if the internal processes do not possess intrinsic symmetry as Joule heating does, i.e. with $\gamma \neq$ 1/2, the result can be quite different in the limit $R/R''\rightarrow\infty$. The efficiency at maximum power in such case is however independent of $\Psi$ as the external dissipation is negligible compared to the internal one.\\

\paragraph*{Additional remarks}
If all dissipation is produced internally, heat is trapped and cannot be extracted efficiently as the thermal conductance under open circuit condition $K_{0}$ is reduced to zero. As a consequence, the internal temperature of the device, possibly quite different from the mean temperature $T'$, may become very high for a macroscopic engine. Preclusion of this unwanted effect is possible with ballistic devices such as that presented by Esposito \cite{Esposito2009}: all the heat is indeed produced at the interfaces, thus avoiding internal warming. This can still be considered as internal dissipation since it is caused by a mesoscopic phenomenon analogous to Joule effect with half of the produced heat released on each side \cite{Gurevich1997}.

As shown in Fig.~(\ref{fig:figure2}) and (\ref{fig:figure3}), the variation of $\eta_{P_{\rm max}}$ for the whole range of $R/R''$ is quite small: while we believe the distinction between internal and external dissipation is of primary importance from a theoretical point of view, it seems of limited interest for technological application.

\paragraph*{Conclusion} Using the example of a thermoelectric generator, a touchstone for irreversible thermodynamics theories \cite{deGroot}, we demonstrated a general result on heat engines: the Curzon-Ahlborn efficiency, though fundamental in the frame of linear irreversible thermodynamics \cite{VandenBroeck2005}, is not a truly universal upper bound on efficiency at maximum power of real heat engines, as inferred in, e.g., Refs.~\cite{Leff87,Gordon89,VandenBroeck2005}, but pertains to endoreversible engines only, whereas the Schmiedl-Seifert efficiency stands only for exoreversible engines, where dissipation is fully internal. The distinction between these two general forms of efficiency at maximum power thus brings a much needed conceptual clarification in finite time thermodynamics. We also showed that the efficiency at maximum power of real heat engines may vary \emph{continuously} between these two extremes, as the sources of irreversibility are tuned. Last but not least, the analysis of the configuration when dissipation contributions are mixed, particularly for dissymetric thermal contacts, raises new questions which are left open.\\

\paragraph{Acknowledgments}
This work is part of the CERES 2 and ISIS projects funded by the Agence Nationale de la Recherche. Y. A. acknowledges financial support from the Minist\`ere de l'Enseignement Sup\'erieur et de la Recherche.

\end{document}